\documentclass{pasj00} 
\usepackage{mathrsfs}  

\SetRunningHead{Y. Sofue and H. Nakanishi}{3D ISM Distribution in the Milky WayIV: Molecular Fraction} 
\Received{2016/mm/dd}  \Accepted{2016/mm/dd}

\title{Three-Dimensional Distribution of the ISM in the Milky Way Galaxy:
IV. 3D Molecular Fraction and Galactic-Scale HI-to-H$_2$ Transition}
 
\author{Yoshiaki \textsc{Sofue} }   
\affil{Insitute of Astronomy, The University of Tokyo, Mitaka, Tokyo 181-0015, Japan, \email{sofue@ioa.s.u-tokyo.ac.jp}}

\author{Hiroyuki \textsc{Nakanishi} }
\affil{Graduate Schools of Science and Engineering, Kagoshima university, 1-21-35 Korimoto, Kagoshima 890-8544, Japan }


\maketitle  
\begin{document} 
 
\def\sun{{_{\odot \hskip-4.9pt \bullet}}}  
\def\Msun{M_{\odot \hskip-4.9pt \bullet}} 
\def\msun{$M_{\odot \hskip-4.9pt \bullet}$} 
\def\kms{km s$^{-1}$}  \def\Vsun{V_0}  \def\Vrot{V_{\rm rot}}
\def\deg{^\circ} 
\def\fmol{f_{\rm mol}}
\def\fmolrho{f_{\rm mol}^\rho}
\def\frho{ $\fmolrho$ }
\def\fmolsigma{f_{\rm mol}^\Sigma} 
\def\fsigma{ $\fmolsigma$ } 
\def\htwo{H$_2$ } 
\def\cc{cm$^{-3}$ }
\def\Z{\mathscr{Z}}
\def\XHI{X_{\rm HI}} \def\XCO{X_{\rm CO}}
\def\IHI{I_{\rm HI}} \def\ICO{I_{\rm CO}}
\def\NHI{N_{\rm H}}  \def\NHH{N_{\rm H_2}} 

\begin{abstract}  
Three dimensional (3D) distribution of the volume-density molecular fraction, defined by $\fmolrho=\rho_{\rm H_2}/(\rho_{\rm HI}+\rho_{\rm H_2})$, is studied in the Milky Way Galaxy. The molecular front appears at galacto-centric distance of $R\sim8$ kpc, where the phase transition from atomic to molecular hydrogen occurs suddenly with $\fmolrho$ dropping from $\sim 0.8$ to 0.2 within a radial interval as narrow as $\sim 0.5$ kpc. The front in density $\fmolrho$ is much sharper than that for surface density molecular fraction.
The front also appears in the vertical direction with a full width of the high-$\fmolrho$ disk to be $\sim 100$ pc.
The radial and vertical $\fmolrho$ profiles, particularly the front behaviors, are well fitted by theoretical curves  calculated using the observed density profile and assumed radiation field and metallicity with exponential gradients. 
The molecular fraction was found to be enhanced along the Perseus and some other spiral arms. The $\fmol$ arms imply that the molecular clouds are produced from HI gas in the spiral arms and are dissociated in the interarm region. We also show that there is a threshold HI density over which the gas is transformed into molecules.
\end{abstract}

\section{INTRODUCTION}
  
The atomic (HI) and molecular (\htwo) hydrogen gases are the major constituents of the interstellar gas. The \htwo\ gas is dominant in the inner Galaxy, while HI in the outer Galaxy. The phase transition from HI to \htwo\ gases is determined by the environmental circumstances in the Galaxy such as the pressure or the density, radiation field and the metallicity. The molecular fraction, the ratio of molecular to total gas densities, is one of the fundamental quantities to represent the ISM conditions (Elmegreen 1993; 
Krumholz et al. 2009). 

Galactic-scale variation of the molecular fraction has been studied in the Galaxy and nearby galaxies using HI and CO line observational data (Sofue et al. 1995; Honma et al 1995; Imamura and Sofue 1997; Hidaka and Sofue 2002; Nakanishi et al. 2006: Nakanishi and Sofue 2016; Tosaki et al. 2011; Tanaka et al. 2014). It was found that the molecular fraction is close to unity in the central regions, and decreases toward the outer galaxy. The transition of the gaseous phases from \htwo\ to HI occurs in a relatively narrow galactocentric region called the molecular front.

In the current studies, observational molecular fraction has been represented in terms of the surface densities (column densities) of HI and H$_2$ gases. In order to compare with the theoretical analyses, it is more appropriate to measure the quantities in terms of volume densities. In this paper we derive the volume-density molecular fraction in the Milky Way Galaxy, and investigate its three-dimensional distribution using the HI and \htwo density cubes obtained by Nakanishi and Sofue (2003, 2006, 2016).

\begin{figure} 
\begin{center}     
\includegraphics[width=7cm]{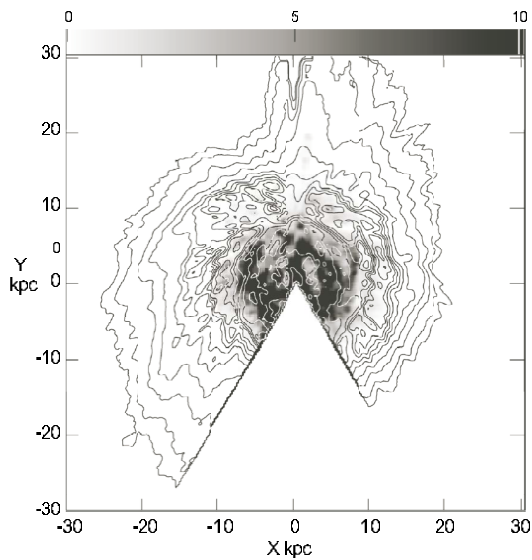}  
\end{center}
\caption{HI (contours at $0.5\times 1$, 2, 5,10, 20, 30, ... K km s$^{-1}$) and \htwo (gray from 0 to 10 K km s$^{-1}$) face-on integrated intensity maps for a region at $-30\le X \le +30$ kpc and $-30 \le Y \le +30$ kpc.}
\label{HIH2face}  
\end{figure}

\begin{figure}
\begin{center}     
\includegraphics[width=7cm]{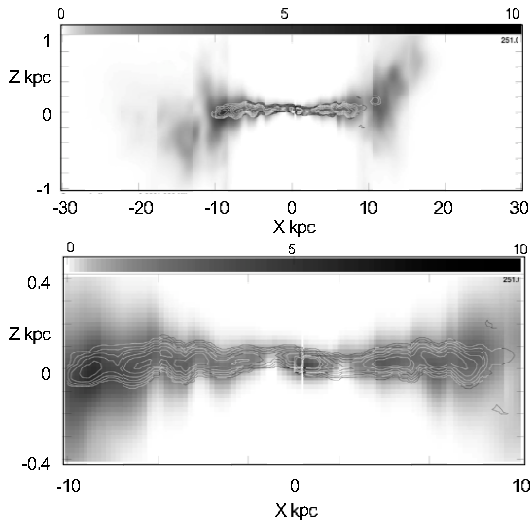}\\ 
\end{center}
\caption{HI (gray) and \htwo (contour) cross section at  $Y=0$  kpc for $-30\le X \le +30$ kpc; $-1 \le Z \le +1$ kpc (top);
and close up for $-10\le X \le +10$ kpc; $-0.4 \le Z \le +0.4$ kpc. HI density is in gray scale as indicated by the bar (0 to 10 H cm$^{-3}$), and molecular density is shown by contours at 1, 2, 4, 8, 16, 32, ... ${\rm H~cm^{-3}} (=2N_{\rm H_2})$.
Note that the $Z$ scale is enlarged by 2.5 times the $X$ axis.
}
\label{HIH2cut}
\end{figure}

\section{3D Molecular Fraction}

\subsection{3D distributions of HI and ${\rm H}_2$ gases}

Figure \ref{HIH2face}  shows the distribution of integrated intensities of HI and CO ($J=1-0$) line emissions ($\IHI$ and $\ICO$, respectively) projected onto the galactic plane (Nakanishi and Sofue 2016). They are proportional to the column densities of neutral hydrogen H and H$_2$ molecules  $\NHI$ and $\NHH$, which are given by $\NHI=\XHI \IHI$ and $\NHH=\XCO \ICO$. 
Here,  $\XHI=1.82\times 10^{18} {\rm cm}^{-2} ({\rm K~km~s}^{-1})^{-1}$  is the conversion factor of the HI intensity to H atop column density. The CO-to-\htwo\ conversion factor has been obtained from various analyses of emprical relations (Bottalo et al. 2013), and is converged to around  $\XCO\simeq 2\times 10^{20}{\rm cm}^{-2} ({\rm K~km~s}^{-1})^{-1}$. In this paper, we adopt  $\XCO=1.8\times 10^{20}{\rm cm}^{-2} ({\rm K~km~s}^{-1})^{-1}$ following Dame et al. (2001).

Figure \ref{HIH2cut} shows meridional cross sections of the gas disk across the Galactic Center showing the volume densities $n_{\rm HI}$ and $n_{\rm H_2}$ in the $(X,Z)$ plane (Nakanishi and Sofue 2016). The molecular gas is distributed in a thin and dense disk near the galactic plane, embedded in the broader and more extended HI disk.

\begin{figure*} 
\begin{center}     
\includegraphics[width=12cm]{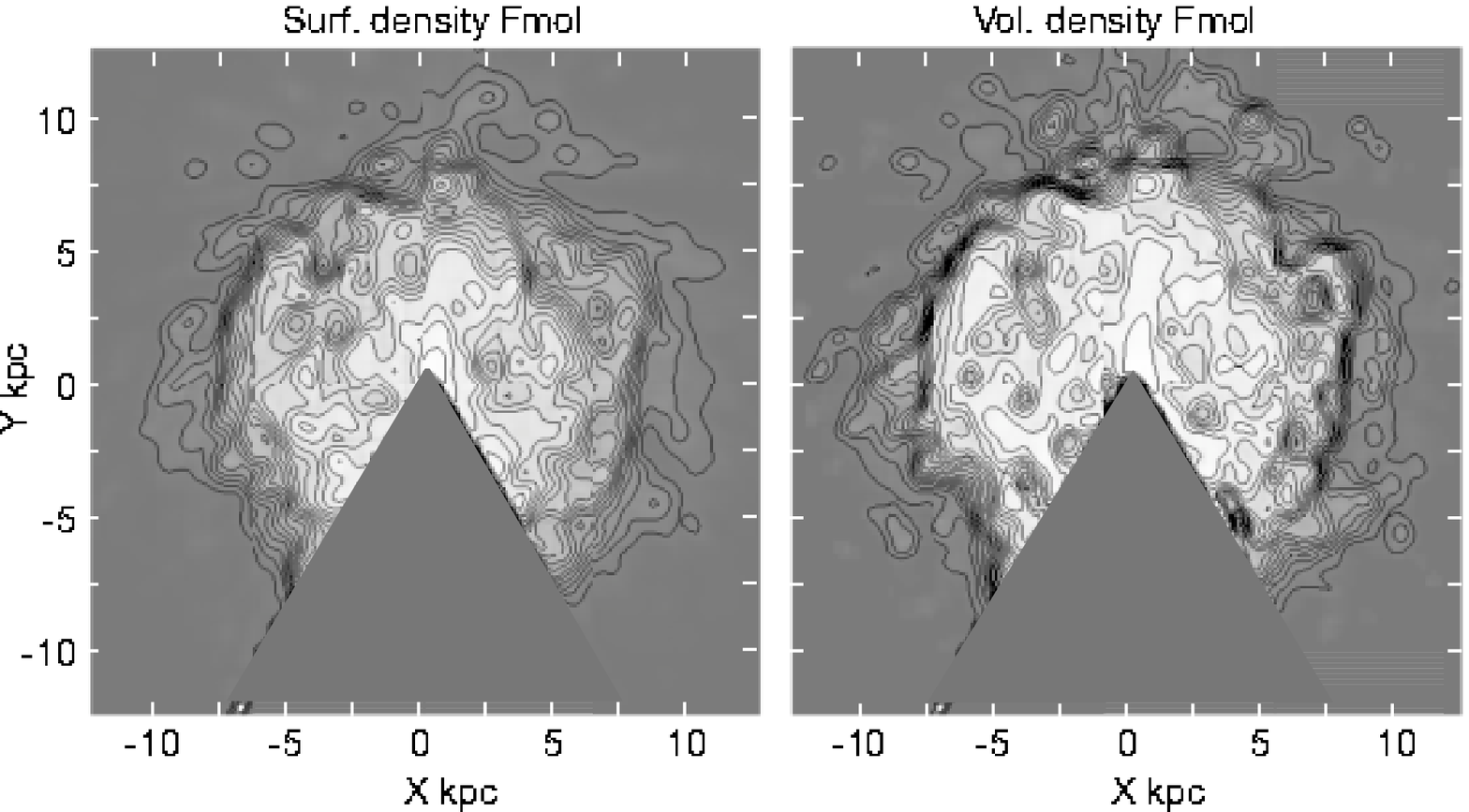}
\end{center} 
\caption{(Left) Surface density molecular fraction $\fmolsigma$ in the $(X,Y)$ plane. Contour interval is 0.1. The contours are drawn at every $\Delta \fmolsigma= 0.1$ starting from 0.1 with according gray scales.  
(Right) Same as left, but for the volume density molecular fraction $\fmolrho$ at $Z=0$ kpc in the $(X,Y)$ plane.  The contours are drawn at every $\Delta \fmolrho= 0.1$ starting from 0.1 with according gray scales. } 
\label{fmol2D} 
\label{fmol3D}  

\begin{center}  
 \includegraphics[width=10cm]{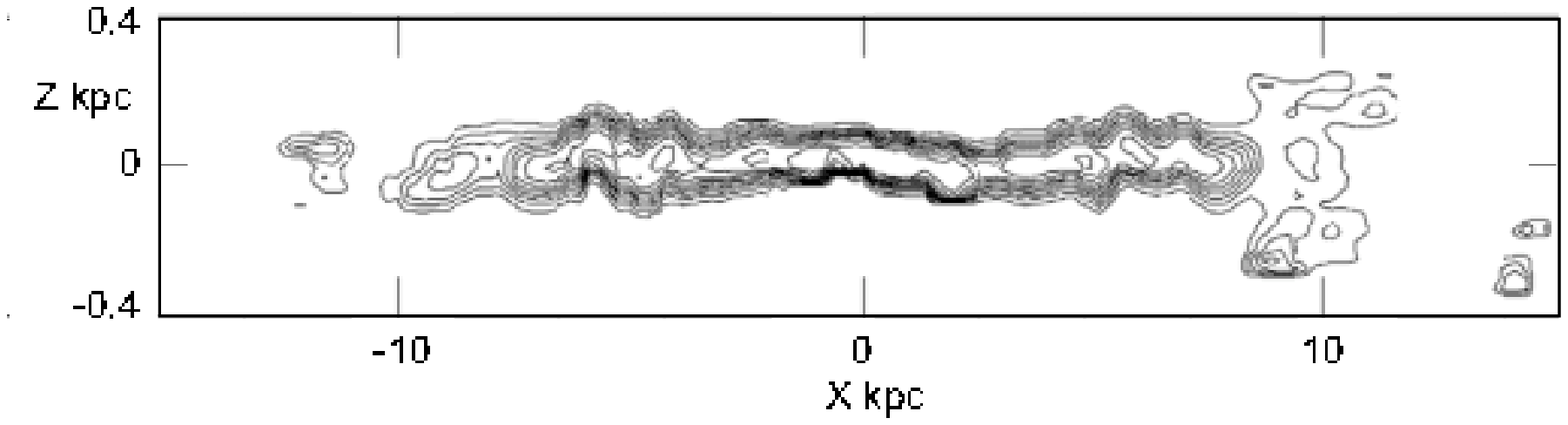}  
 \end{center}  
\caption{ $(X,Z)$ cross section of the 3D density molecular fraction $\fmolrho$ at $Y=0$.  Contours are drawn at every $\Delta \fmolrho= 0.1$ starting from 0.1.}  
\label{fmolxzy}

  \begin{center}   
  \includegraphics[width=12cm]{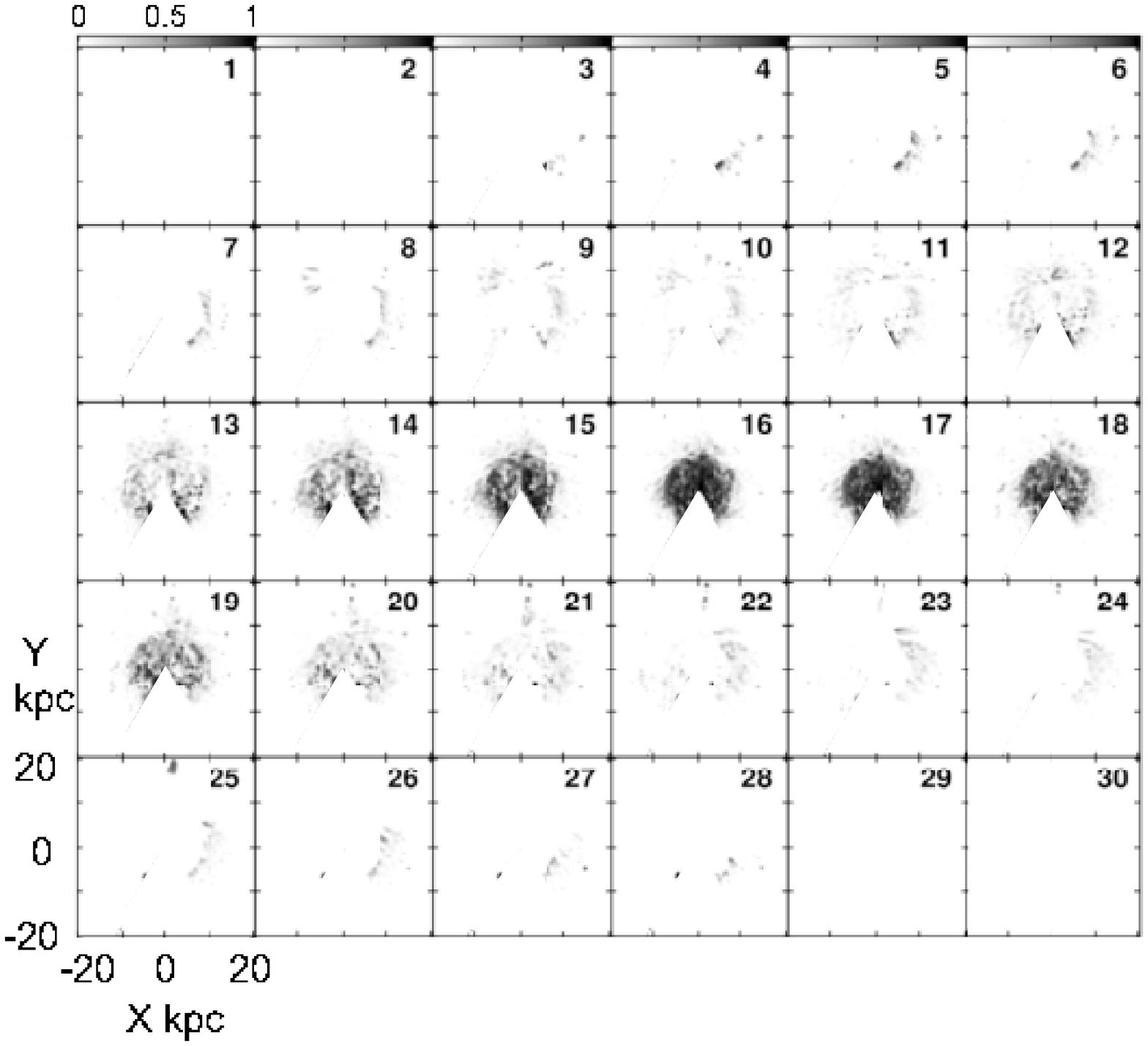}  
\caption{Density molecular fraction $\fmolrho$ at different heights $Z$ from the galactic plane at 40 pc $Z$ increment. The values are represented in gray scale as indicated by the bar in the top-left corner. }
\label{fmolxyz}
\end{center}
\end{figure*}

\subsection{3D molecular fraction}

The surface-density (column density) molecular fraction is defined by 
\begin{equation}
\fmolsigma={\Sigma_{\rm H_2} \over {\Sigma_{\rm HI} + \Sigma_{\rm H_2}}}
={{2N_{\rm H_2} \over {N_{\rm HI}} + 2N_{\rm H_2}}},
\label{fmolsigma}
\end{equation}
where $\Sigma_{\rm HI}$ and $\Sigma_{\rm H_2}$ are the surface mass densities of neutral hydrogen and molecular gases, respectively, integrated in the direction perpendicular to the galactic plane.

The volume-density molecular fraction is defined by 
\begin{equation}
\fmolrho={\rho_{\rm H_2} \over {\rho_{\rm HI} + \rho_{\rm H_2}}}
={2n_{\rm H_2} \over n_{\rm HI}+ 2  n_{\rm H_2}},
\label{fmolrho}
\end{equation}
where $\rho_{\rm HI}$ and $\rho_{\rm H_2}$ are the volume densities of the HI and molecular hydrogen gases, and $n_{\rm HI}$ and $n_{\rm H_2}$ are the volume number densities, respectively.

In our previous paper (Nakanishi and Sofue 2016), we constructed HI and \htwo\ density cubes by applying the velocity-to-space transformation to the kinematical data (line brightness temperature, LSR velocity, galactic longitude and latitude) of HI- and CO-line spectral data using a galactic rotation curve assuming circular rotation of the Galaxy. Thereby, we used the Bonn-Leiden-Argentine All Sky HI Survey (Karlbera et al. 2005) and the Colombia Galactic Plane CO-Line Survey (Dame et al. 2001). Using the thus obtained density cubes, we constructed a three dimensional cube of the surface- and volume-density molecular fractions, $\fmolsigma$ and $\fmolrho$, in the entire Galaxy.  

Figure \ref{fmol2D} shows two dimensional distributions of the surface-density molecular fraction $\fmolsigma$, and figure \ref{fmol3D} shows volume-density molecular fraction $\fmolrho$ in the galactic plane at $Z=0$. 
Figure \ref{fmolxzy} shows vertical cross section of the 3D \frho map at $Y=0$.
Figure \ref{fmolxyz} shows sliced maps of $\fmolrho$ distribution at different heights at every $\Delta Z=40$ pc from $Z=-500$ pc to $+500$ pc. 

Both figures  \ref{fmol2D} and \ref{fmol3D} show that the high-molecular fraction region is located in the inner Galaxy at $R<\sim 8$ kpc, but the volume-density molecular fraction shows much clearer plataeu-like distribution. The edge of the plateau is called the molecular front, where the galactic scale phase transition from HI to H$_2$ is occuring.

\subsection{Radial molecular front}

\begin{figure} 
\begin{center}  
 \includegraphics[width=7cm]{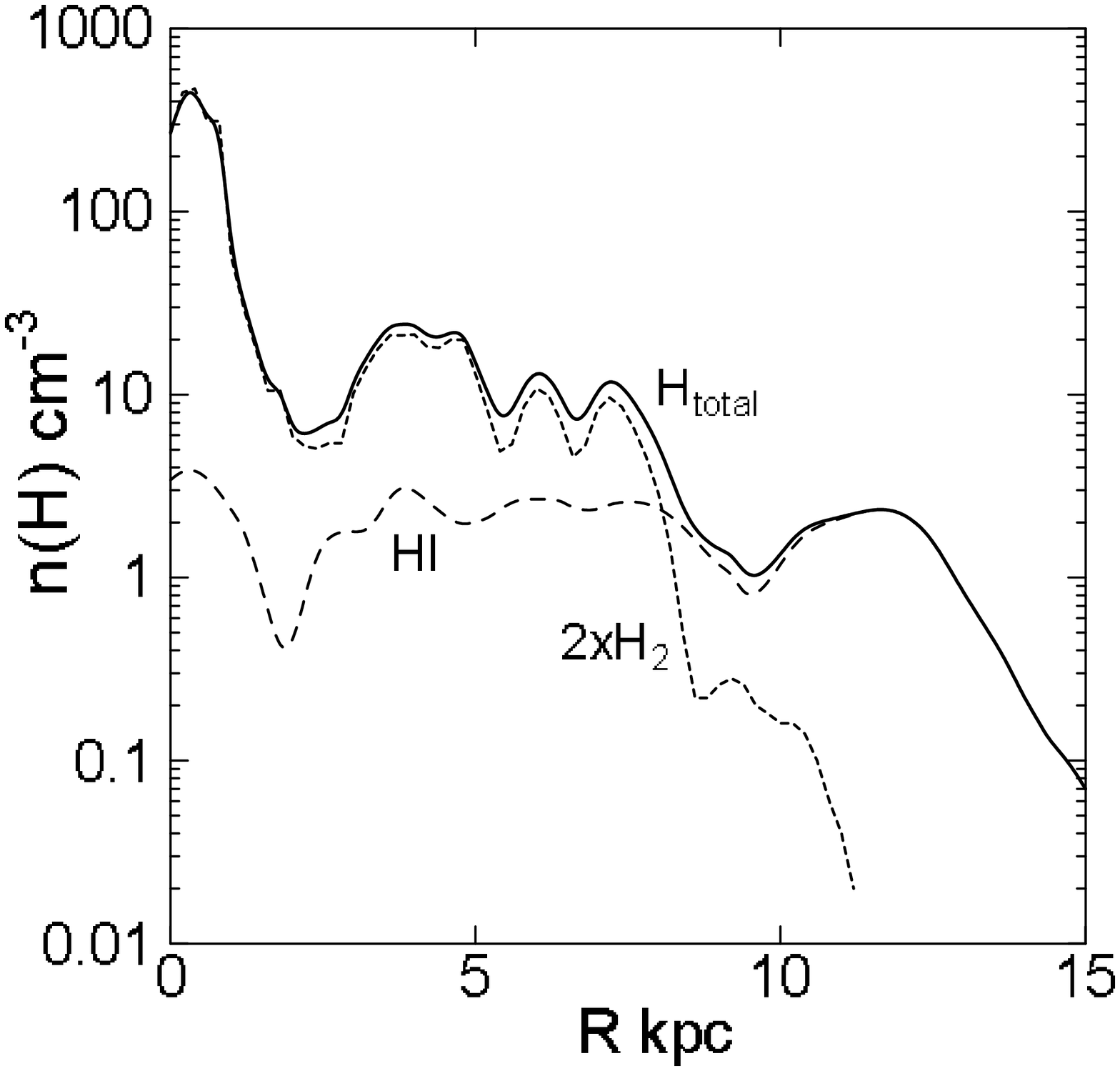}    
\end{center}
\vskip -4mm\caption{Volume density $n({\rm H})$ ($=n_{\rm HI}$, $2n_{\rm H_2}$, or $n_{\rm tot}$) plotted against $R$ in the galactic plane along the $X$ axis. }
\label{r_Htot} 

\begin{center}   
 \includegraphics[width=7cm]{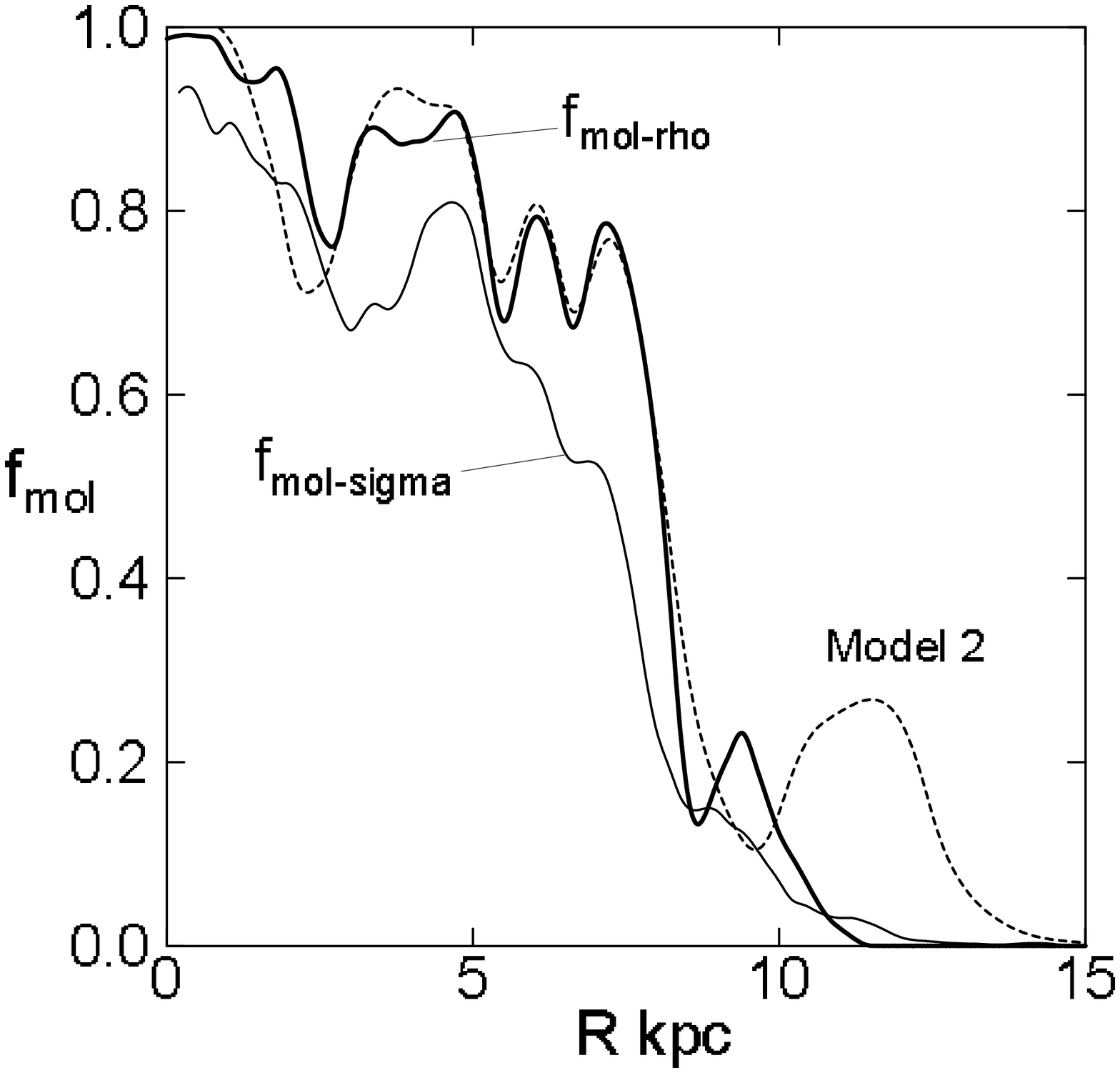} 
\end{center}
\vskip -2mm\caption{ \frho\ (thick line) compared with $\fmolsigma$ (thin line) plotted against $R$. The dashed line shows a calculated curve based on the Elmegreen's (1993) model. }
\label{r_fmol} 

\begin{center}    
\includegraphics[width=7cm]{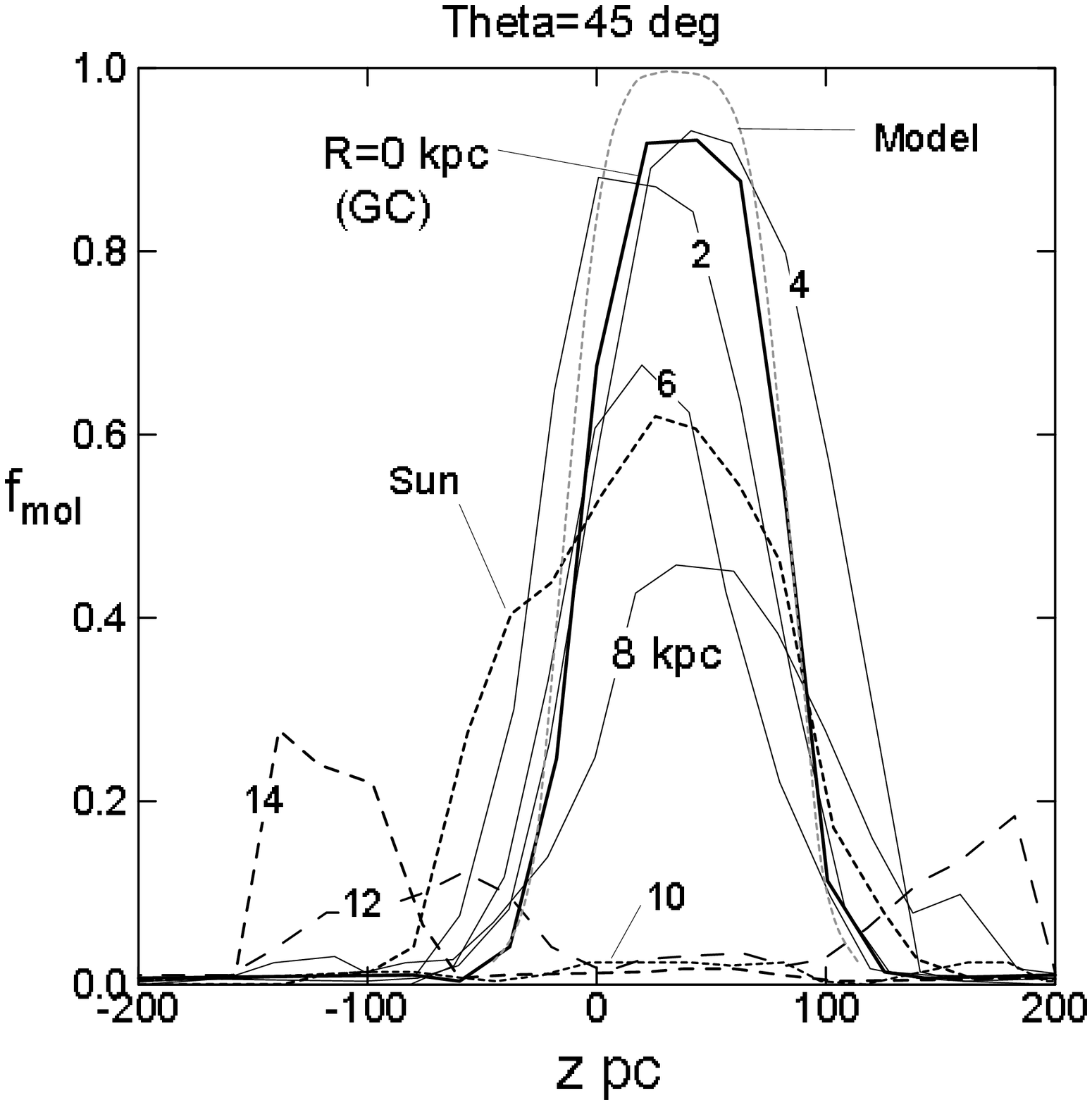} 
\end{center}
\vskip -2mm\caption{Vertical variation of \frho\ at $R=0$ kpc (Galactic Center), 2, 4, 6, 8 and Sun, 10, 12, and 14 kpc at a galacto-centric longitude $\Theta=45^\circ$. The gray dotted line is a model curve.
 The vertical variation near the Sun ($R=8$ kpc, $\Theta=90^\circ$ is also indicated by the thick dashed line.
 } 
\label{z_fmol}
\end{figure}

Figure \ref{r_fmol} shows radial variations of the volume densities of HI ($n_{\rm HI}$), \htwo\ ($2n_{\rm H_2}$), and total gas density ($n_{\rm tot}=n_{\rm HI}+2n_{\rm H_2}$) in the galactic plane as a function of the galacto-centric distance along the $X$ axis ($Y=0)$. In the figure, $n({\rm H})$ represents either $n_{\rm HI}$, $2n_{\rm H_2}$, or $n_{\rm tot}$.  
The figure also shows the corresponding volume-density molecular fraction $\fmolrho$. It shows also the surface-density molecular fraction $\fmolsigma$ in the $(X,Z)$ plane calculated for the integrated column densities in $Z$ direction. 

In the present analysis, we used densities along the $X$ axis at $Y=0$ kpc, where kinematical distances, and therefore densities, are determined in better accuracy than along the $Y$ axis across the Sun.
 
The volume-density molecular fraction is as high as $\fmolrho\sim 0.9$ at $R\le 5$ kpc, and steeply decreases to low values less than 0.1 at $R>9$ kpc with an $e$-folding radius interval of $\Delta R\sim 0.5$ kpc from $R=7.5$ to 8 kpc. 
The figure indicates that the high molecular fraction region composes a plateau-like disk with critical radius  $R^\rho_{\rm F}\simeq 8$ kpc, where the fraction decreases sharply with radius, which we call the front radii. 

On the other hand, the surface-density molecular fraction varies smoothly with the maximum value of about $\fmolsigma\sim 0.7$ at $R\le 5$ kpc. Such a mild variation of $\fmolsigma$ is due to averaging effect of the thicker HI disk having smaller molecular fraction at high $Z$ regions as seen in figure \ref{HIH2cut}.

The distribution of $\fmolsigma$ may be compared with those obtained for external galaxies: Similar front structure has been observed in nearby spiral galaxies (Sofue et al. 1995; Honma et al. 1997; Tanaka et al. 2014). A milder variation has been observed in later type galaxies like M33 (Tosaki et al 2011), where the maximum fraction was found to be about $\fmolsigma \simeq 0.22$ at $R\sim 0.6$ kpc. 

It should be emphasized that the volume-density molecular fraction obtained here in the Milky Way shows a much clearer, sharper front behavior than those ever observed for surface molecular fraction in galaxies.
 
\subsection{Vertical molecular front}

The 3D distribution of the molecular fraction makes it possible to examine the vertical variation of $\fmolrho$ in details. Figure \ref{fmolxzy} shows a cross section of the cube in $(X,Z)$ plane across the Galactic Center. 
Figure \ref{z_fmol} shows vertical variations of $\fmolrho$ at different radii from $R=0$ kpc at the Galactic Center to $R=14$ kpc at a galacto-centric longitude $\Theta=45\deg$. This figure indicates that the galactic disk inside the solar circle, particularly at $\le 5$ kpc, is molecular-gas dominant with $\fmolrho \sim 0.8-0.93$ and as high as $\sim 0.5$ even near the solar circle. It then drastically decreases outside the solar circle.  

It should be stressed that the $\fmolrho$ decreases sharply with the height at $\pm 50$ pc from the midplane of maximum $\fmolrho$, exhibiting vertical molecular front of the molecular disk at $h_{\rm c}\simeq \pm 50$ pc. The shapes of the vertical and radial molecular fronts are similar except for the scale radius and height.

\begin{figure} 
\begin{center}     
\includegraphics[width=8cm]{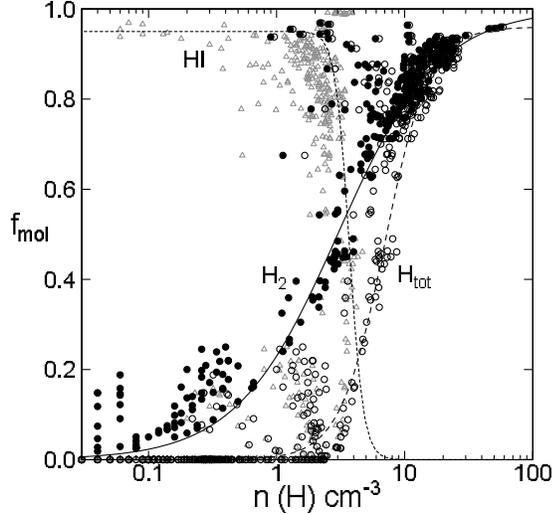} \\
\end{center} 
\caption{$\fmolrho$ plotted against densities of HI $n({\rm HI})$ (gray triangle), molecular hydrogen $2n({\rm H_2})$ (full circle), and total atomic hydrogen $n_{\rm total}=n({\rm HI})+2n({\rm H_2})$ (open circle). The lines are approximate empirical fits as described in the text.}
\label{rho_fmol} 
\end{figure}

\subsection{Dependence of \frho on the ISM density: Empirical law and HI threshold}

\def\a{\alpha}

It is known that the molecular gas is formed from and dissociated into HI gas responding to the interstellar condition. The molecular fraction is a function of the gas density through the pressure $P=n_{\rm tot}kT$, metallicity $\Z$, and UV radiation field intensity $j$ (Elmegreen 1993), where $n_{\rm tot}=n_{\rm HI}+2n_{\rm H_2}$, $T$ is the equivalent ISM temperature, and $k$ is the Boltzman constant.

We now examine the dependence of \frho on the densities, $n({\rm H})$. Figure \ref{rho_fmol} plots $\fmolrho$ against $n({\rm H})$, where $n({\rm H})$ represents either $n_{\rm HI}$, $2n_{\rm H_2}$ or $n_{\rm tot}$. 

The dependence of the molecular fraction on the gas densities can be represented by an empirical simple function as 
\begin{equation}
\fmolrho={f_{\rm max}\over 1+(n/n_{\rm c})^{\a}},
\label{fmolfunc}
\end{equation}
where $n_{\rm c}$ is a critical density depending on the species.

This empirical law for total gas density is approximated by a parameter combination 
$f_{\rm max}=0.96$, $n_{\rm c}=7.0~{\rm H~ cm^{-3}}$, and $\a=-2.5$,
 as shown by open circles in figure \ref{rho_fmol} and fitted by the dashed curve. The critical density of 7 H cm$^{-3}$ is close to that of the molecular front in figure \ref{r_fmol}. The value may represent a threshold ISM density to generate the cold neutral medium (CNM) from warm (WNM) phase  
 (Heiles and Troland 2003).
 
Similarly, \frho depends on the molecular gas density by 
$f_{\rm max}=1.0$, $n_{\rm c}=3.0~{\rm H~ cm^{-3}}$ and $\a=-1.1$,
as shown by the filled circles fitted by the thin full line. 
Table \ref{rhofmolfit} summarizes the parameters.

On the other hand, the dependence on the HI density (gray triangles in figure \ref{rho_fmol}) is quite different. The  dotted curve in figure  \ref{rho_fmol} shows the same function for parameters
$f_{\rm max}=0.95$, $n_{\rm c}=3.7~{\rm H~ cm^{-3}}$, and $\a=+6$, representing a sharply dropping \frho near an HI threshold value of $n_{\rm HI}\simeq 3.7~{\rm H~cm}^{-3}$. This implies that there is a maximum HI density, exceeding which the HI gas is transformed into molecular gas.  

\begin{table}
\caption{Parameters for the empirical fit to  $n$-\frho plot in figure \ref{rho_fmol} by $\fmolrho=f_{\rm max}/[1+(n/n_{\rm c})^{\a}]$. }
\begin{center}
\begin{tabular}{llll}  
\hline
Phase & $f_{\rm max}$ & $n_{\rm c} ({\rm H~ cm^{-3}})$ & $\a$ \\ 
\hline
HI & 0.95 & 3.7 & $+6$ \\  
H$_2$ &1.0 &3.0 & $-1.1$ \\ 
H$_{\rm tot}$ & 0.96 & 7.0 &$-2.5$ \\ 
\hline
\end{tabular}
\end{center}
\label{rhofmolfit}
\end{table} 

\begin{figure} 
\begin{center}    
\includegraphics[width=7cm]{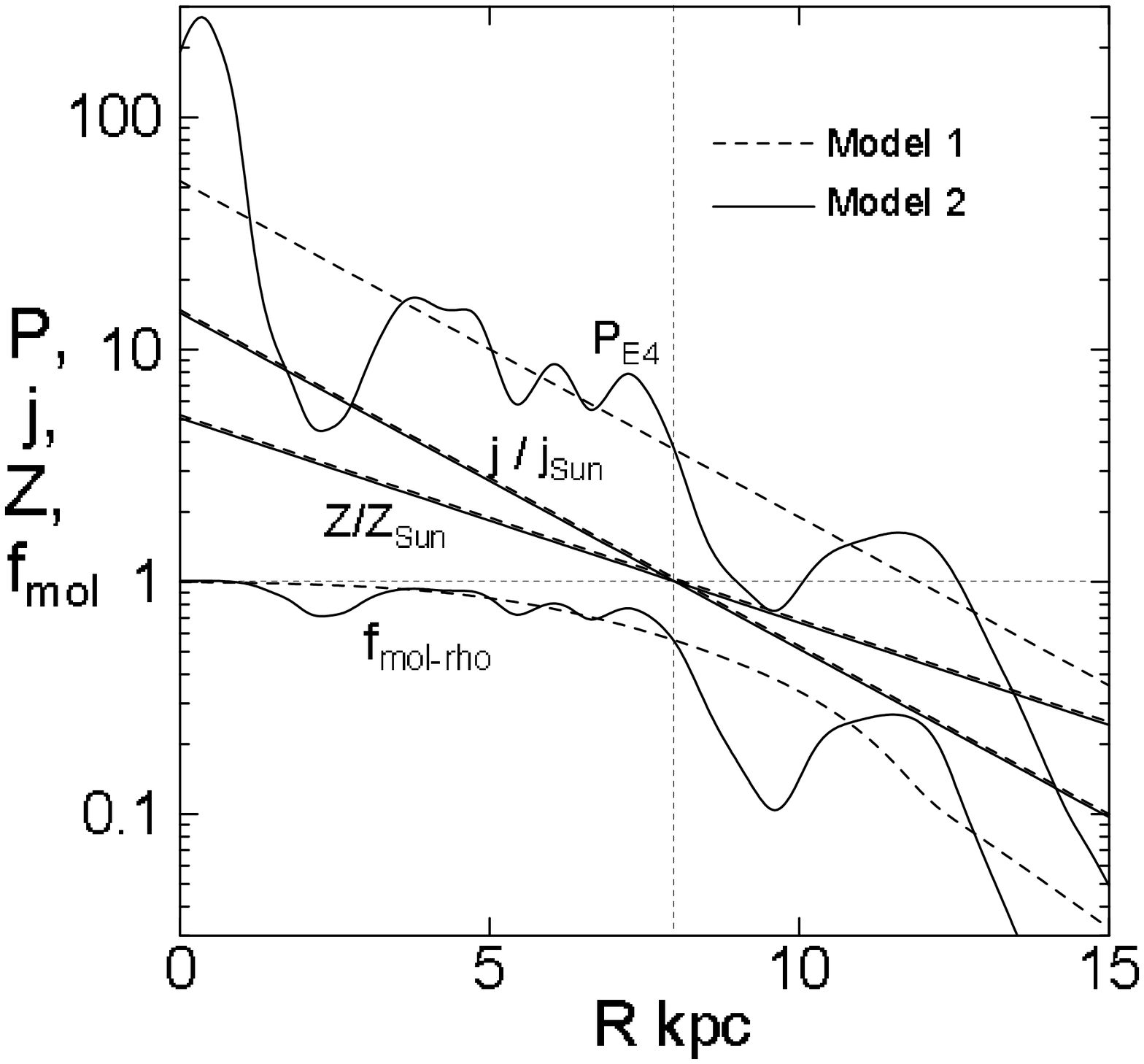}  \\   
\includegraphics[width=7cm]{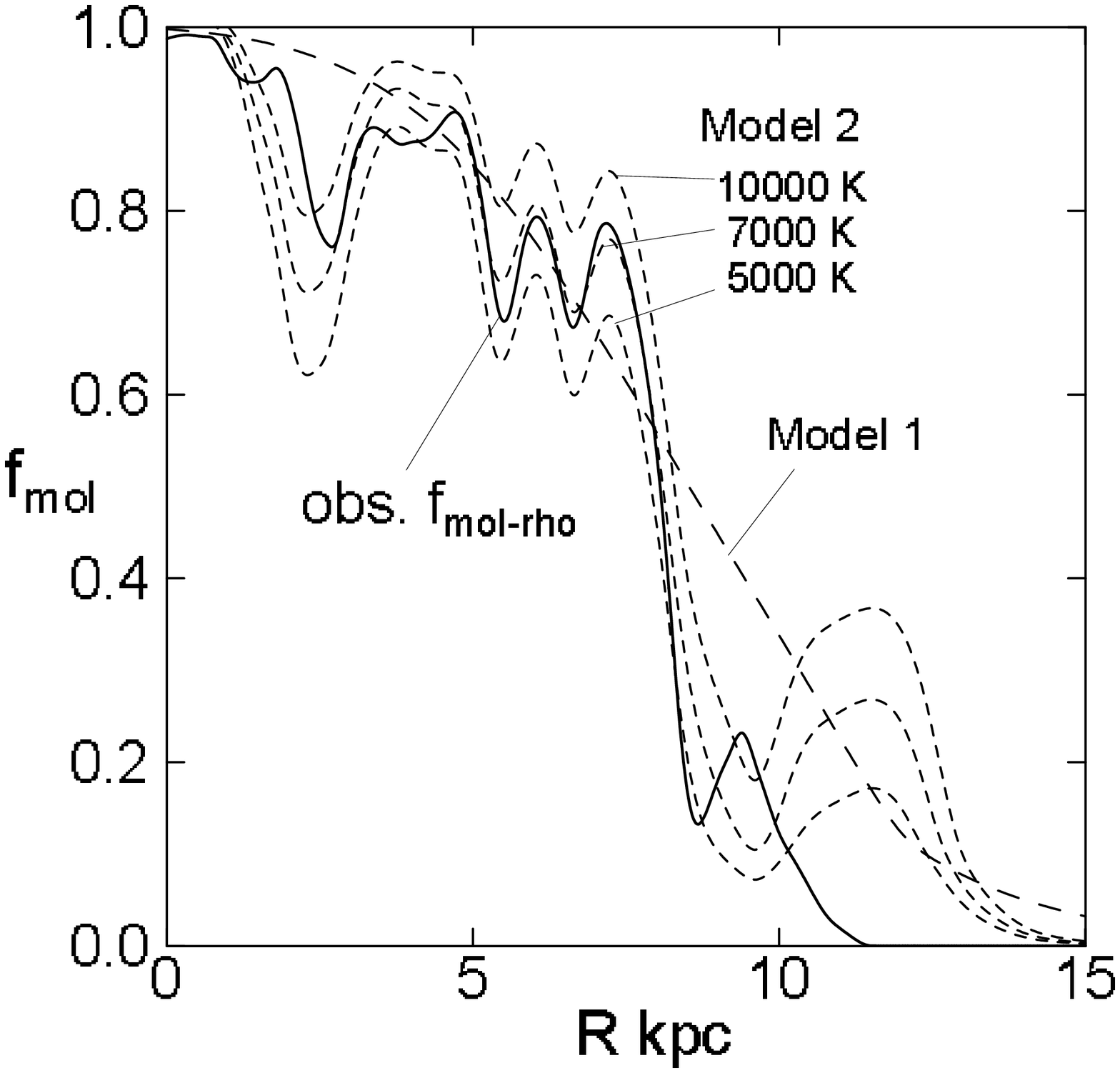}\\ 
\end{center}
\caption{(Top) Radial variations of $P_{\rm E4}=P/(10^4 {\rm K~cm^{-3}})$, $j/j_0$, $\Z/\Z_0$ and \frho for Model 1 and 2 for $ T=7000$ K.
(Bottom) Comparison of \frho models with the observation (thick line): Model 1 for ISM temperature  7000 K (long dash), and Model 2 for 5000, 7000 and 10000 K (dashed lines).}
\label{r_fmol_comparison} 
\end{figure}  

\section{Comparison with Theoretical Models} 

The variation of $\fmolrho$ in the $R$ and $Z$ directions can be calculated by a theoretical consideration of the phase transition in the interstellar gas taking account of the ISM pressure $P$, metallicity $\Z$, and ultra-violet radiation (UV) field by stars $j$ (Elmegreen 1993; Tanaka et al. 2014). 

We consider two cases of pressure distribution. In the first model (Model 1), the pressure $P$ is expressed by an exponential function, and in the second semi-analytical model (Model 2) we adopt the observed total density distribution along the $X$ axis at $Z=0$ and $Y=0$ kpc as shown in figure \ref{r_Htot}. This direction was chosen for accurate densities than those along $Y$ axis (Sun-Galactic Center line), where the density is less certain for degenerated kinematic distances.

In Model 1 $P$ is given by 
\begin{equation}
         P = P_0 e^{-(R-R_0)/R_P},
\end{equation}
where the scale radius is taken to be $R_P=3$ kpc. 
In Model 2, the pressure $P$ is assumed to be proportional to the observed total hydrogen density as
\begin{equation}
P=n_{\rm tot} k T.
\end{equation} 
We normalize the density by $n_{\rm tot:0}=5.3$ \cc from the present measurement at $R=X=8$ kpc as an approximate representative value at the Sun. The temperature is assumed to be constant at $T=7000$ K, equivalent to the velocity dispersion of clouds and interstellar turbulence, which is close to the pressure-equilibrium kinematic temperature, 8000 K, of the warm neutral medium (Heiles and Troland 2003).  Here and hereafter, the suffix 0 denotes the values at $R=8$ kpc approximately representing the solar value.

In both models, we adopt the same profiles for $j$ and $\Z$ as
\begin{equation}
         j =j_0 e^{-(R-R_0)/R_j}
\end{equation}
and
\begin{equation}
         \Z = \Z_0 e^{-(R-R_0)/R_Z}.
         \label{ZR}
\end{equation}
The radiation field is assumed to have the same scale radius as the pressure, $R_j=3$ kpc, and the solar value $j_0=0.8\times 10^{-12}$ erg cm$^{-3}$ (Cox 2000) is the same as taken by Elmegreen (1993).

The metallicity gradient was determined by Shaver et al. (1983), who obtained
$ d\log \Z/dR =(-0.07\pm 0.015)$ dex kpc$^{-1}$, where they assumed $R_0=10$ kpc.
By rescaling the radius for $R_0=8$ kpc, the gradient can be rewritten as 
$ d\log \Z/dR =(-0.7\pm 0.15)/(8~{\rm kpc})$, or
$d{\rm ln}\Z/dR= -0.20\pm 0.04~{\rm kpc}^{-1}\simeq 1/(5\pm 1  ~{\rm kpc}) $. 
This yields $R_Z\simeq 5$ kpc, and we adopt this value for equation \ref{ZR}. 

Calculated results for the two models using the code developed by Tanaka et al. (2014) are shown in figures \ref{r_fmol} and \ref{r_fmol_comparison}.  
The exponential model (Model 1) shows smooth and mild variation. Although it roughly approximates the general behavior of observed \frho, the sharp molecular front cannot be reproduced. On the other hand, the semi-analytical model (Model 2) using the observed density profile well reproduces the observed \frho, when the ISM temperature is taken to be 7000 K. The model can reproduce not only the steep molecular front at $R\sim 8$ kpc but also the fluctuations.

The model profile is sensitive to the ISM temperature. If the temperature is changed to $\sim 10^4$ K or to $\sim 5000$ K, the fitting gets worse, with \frho being increased or decreased significantly, as shown in figure \ref{r_fmol_comparison}(c). This implies that the \frho fitting can be used to determine the ISM temperature, if the radiation field and metallicity are given.

The $Z$ directional density profile is expressed by a hyperbolic cosecant function of the height $Z$. The metallicity and the radiation intensity field are expressed by an exponentially decreasing function with $R$ and $Z$.
The $Z$-directional variation of model $\fmolrho$ is shown in figure \ref{z_fmol} for the inner region of the Galaxy. 
The steep vertical variation of \frho near the vertical molecular front is well reproduced by this model.

\begin{figure*} 
\begin{center}      
\includegraphics[width=14cm]{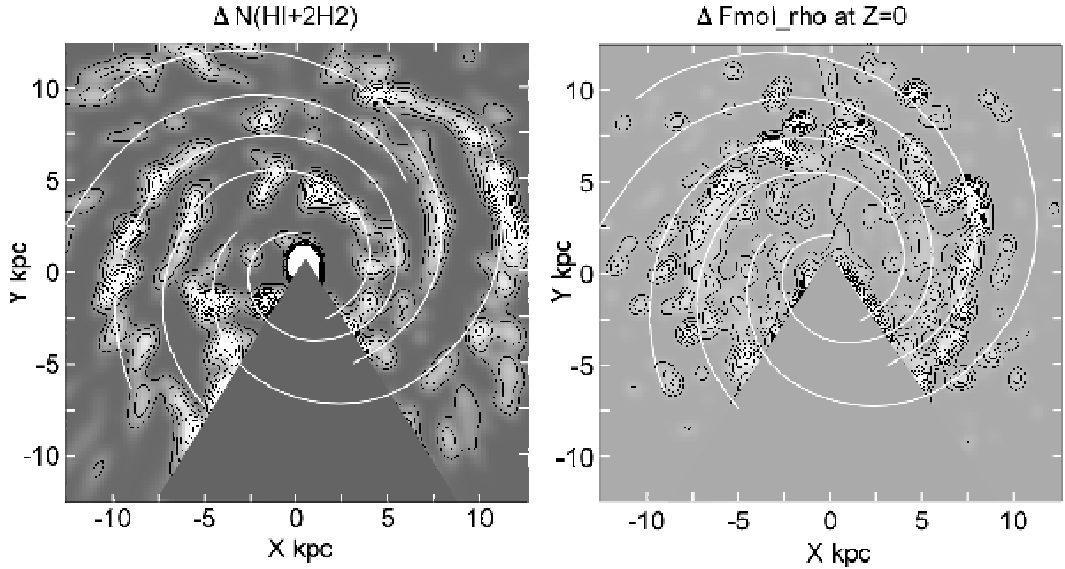} 
\end{center} 
\caption{ 
Unsharp-masked total surface density $\Delta N$ (HI+2H$_2$) (left) and $\Delta \fmolrho$ (right) maps. Contour interval for molecular fraction is 0.02.  Spiral arms traced by Nakanishi and Sofue (2016) are indicated by white lines.  }
\label{deltaFmol} 
\end{figure*}

\section{Spiral Arms}

In order to see the variation of molecular fraction in relation to the spiral arms, we made unsharp-masked map of the density molecular fraction in the galactic plane. 
The original map was smoothed by a Gaussian beam with full width of 5 pixels, or 1 kpc, and the smoothed map was subtracted from the original to yield a difference map, $\Delta \fmolrho$.
The thus obtained difference (unsharp-masked) map is shown in figure \ref{deltaFmol}, displaying only positive $\Delta \fmolrho$ regions. It is also compared with an unsharp-masked map of the total gas density for the same masking parameters. The spiral arms traced in our earlier works are superposed by white lines (Nakanishi and Sofue 2006, 2016). 

We stress that the $\Delta \fmolrho$ map exhibits spiral arms. A clear coincidence is found with the Perseus Arm, where the \frho\ arm almost perfectly traces the spiral arm. 
The $\Delta \fmolrho$ arms are found mainly at radii $R\sim 6$ to 10 kpc near the molecular front. However, they are hardly seen in the innermost region at $R<\sim 5$ kpc, where the molecular fraction is saturated close to unity, and in the outer Galaxy at $R>\sim 10$ kpc, where the gas is almost totally HI. 
 
Hence, in the transition region at  $R=6$ to 10 kpc, molecular clouds are formed within the spiral arms, and are dissociated in the interarm region. 
 This is consistent with the long arm-interarm crossing time  of the spiral pattern,
\begin{equation}
t\sim \pi  /(\Omega -\Omega_{\rm p}),
\end{equation}
which is on the order of $\sim$ several$\times 10^8$ to $10^9$ yrs and is sufficiently longer than the molecular gas formation time scale $t_{\rm mol}\sim 10^7$ yrs (Goldsmith et al. 2007). Here, $\Omega=V_{\rm rot}/R\sim 20 {\rm km ~cm^{-1} kpc^{-1}} $ is the angular velocity of the disk gas for $\Vrot=220$ \kms and $R=8-10$ kpc, and $\Omega_{\rm p}\sim 23{\rm km ~cm^{-1} kpc^{-1}}$ (Junqueira et al. 2015) is the pattern speed.

Besides these \frho\ arms, there is a particular region near$(X,Y)\sim (-5,+5)$ kpc between the Scutum-Crux and Sagittarius-Carina Arms, where \frho\ is enhanced in the interarm region. However, this is due to the sharp edge of the molecular front in this area which causes an artifact of arm-like enhancement when the unsharp-masking was applied.

\section{Summary and Discussion}

\subsection{3D sharp molecular fronts}
We constructed a 3D cube of the volume-density molecular fraction, \frho, using the HI and \htwo\ density cubes of the Milky Way. The radial molecular front in the galactic plane observed in the density molecular fraction \frho is found to be sharper than that for the surface density molecular fraction \fsigma. 
The molecular front  of \frho\ is also found in the vertical direction.

Plots of \frho against HI, \htwo, and total gas densities showed that \frho is a function of gas densities empirically expressed by equation \ref{fmolfunc}. 
A threshold value was obtained in the HI density at $n_{\rm HI}\simeq 3.7 {\rm H~cm}^{-3}$, above which the gas is transformed almost totally into molecular gas.  

\subsection{Reproduction by theoretical models}
The radial and $Z$-directional variations of the observed \frho\ and sharp molecular fronts are well reproduced by the theoretical calculations using the Elmegreen's (1999) model. We showed that the radial profile is a sensitive function of the ISM temperature, which was here taken to be $7000$ K for reasonable fitting. The temperature is consistent with the pressure-equilibrium kinetic temperature, 8000 K, derived for the warm neutral ISM (Heiles and Troland 2003). 

The radial \frho\ profile was analyzed only along $+Y$ axis because of the highest quality of data in this direction as indicated from figures \ref{HIH2face} and \ref{fmol3D}. The profile may represent the typical radial variation in the whole galaxy, and analyses along other radial directions will give almost the same result. Also, azimuthal averaging was not applied, as it smears out the molecular fronts to be studied in this paper due to the globally asymmetric distribution of the gases (figure \ref{HIH2face}).

\subsection{Enhanced-$\Delta \fmolrho$ spiral arms}
 
Two dimensional unsharp-masked $\Delta \fmolrho$ map in the galactic plane exhibited spiral arms with enhanced molecular fraction at $R\sim 6$ to 10 kpc near the molecular front. 
A clear \frho\ arm was found coinciding with the Perseus arm. This fact implies that HI gas is transformed into molecular gas along a spiral arm with marginal \frho near the molecular front (transition region). However, no clear \frho\ arms are seen at $R<6$ kpc or at $R>10$ kpc, where the gas is either totally molecular or HI. In extragalactic systems, an enhancement of $\fmolsigma$ has been observed in outer spiral arms of the Sb galaxy M51 (Hidaka and Sofue 2002).  

\subsection{Effects of cold/opaque HI and dark H$_2$ gases}
 
If we consider cold HI gas, it may increase the total HI density by a factor of $\sim 2$ (Fukui et al. 2015). Moreover, if we considr the finite optical thickness, the HI densities, which have been calculated for optically thin assumption, will be increased. By increasing HI density, $\fmol$ will generally decrease, and the molecular front will be shifted toward inside.   

Simillary, CO-free "dark" \htwo\ gas has been suggested (Wolfire et al. 2010). If such dark gas exists, \frho\ increases and the molecular front will move outward. However, if the gas coexists with molecular clouds, the effect is already included in the conversion factor derived by Virial-mass and $\gamma$-ray mehtods (Bottalo et al. 2013). Then, dark \htwo\ does not affect the result. 

Detailed discussion of the effects of cold/opaque HI and dark molecular gases is beyond the scope of this paper, and will be subject for future works.

\end{document}